\documentclass[aps,prb,twocolumn,superscriptaddress,showpacs]{revtex4-1}
\usepackage{graphicx,bm,color,float,textcomp,mathtools,times}
\newcommand\sho{$\rm SrHo_2O_4$}
\newcommand\bdo{$\rm BaDy_2O_4$}
\newcommand\seo{$\rm SrEr_2O_4$}
\newcommand\sdo{$\rm SrDy_2O_4$}
\newcommand\sRo{SrRE$_2$O$_4$}
\newcommand\afm{antiferromagnetic}

\newcommand\KO{${\bf k} = 0$}
\newcommand\Kone{${\bf k}_1 \! = \! [\frac{1}{2} 0 \frac{1}{2}]$}
\newcommand\Ktwo{${\bf k}_2 \! = \! [\frac{1}{2} \frac{1}{2} \frac{1}{2}]$}

\begin{document}
\title{Fragile ground state and rigid field-induced structures in zigzag ladder compound $\mathbf{BaDy_2O_4}$}
\author{D.D.~Khalyavin} \affiliation{ISIS Facility, STFC Rutherford Appleton Laboratory, Chilton, Didcot, OX11 0QX, United Kingdom}
\author{P.~Manuel} \affiliation{ISIS Facility, STFC Rutherford Appleton Laboratory, Chilton, Didcot, OX11 0QX, United Kingdom}
\author{M.~Ciomaga Hatnean} \affiliation{Department of Physics, University of Warwick, Coventry CV4 7AL, United Kingdom}
\author{O.~A.~Petrenko} \affiliation{Department of Physics, University of Warwick, Coventry CV4 7AL, United Kingdom}
\date{\today}
\begin{abstract}
We report on a sequence of field-induced transitions in the zigzag ladder compound \bdo\ studied with powder neutron diffraction and magnetisation measurements.
In agreement with the previously published results, the low temperature zero-field structure is characterised by two half-integer propagation vectors, ${\bf k}_1=[\frac{1}{2}~0~\frac{1}{2}]$ and ${\bf k}_2=[\frac{1}{2}~\frac{1}{2}~\frac{1}{2}]$. 
However, on application of an external magnetic field, the Bragg peaks corresponding to the zero-field structure lose their intensity rather rapidly and disappear completely in a field of 2.5~kOe. 
In the intermediate fields, 2.5 to 22.5~kOe, new peaks are observed characterised by the propagation vector ${\bf k}^\prime =[0~0~\frac{1}{3}]$ corresponding to an {\it up-up-down} (uud) structure as well ${\bf k}=0$ ferromagnetic peaks.
This regime of fields corresponds to a pronounced plateau in the magnetisation curve. 
Remarkably, the uud structure survives heating to at least 1.4~K, three times higher temperature than the $T_{\rm N}$ of 0.48~K for the zero-field structure.
Above 22.5~kOe, the ${\bf k}^\prime$ peaks disappear while the ${\bf k}=0$ peaks gain significant intensity indicating an increase in the polarisation of the system.
The analysis of the intensities of the field-induced reflections allows for a clear identification of the magnetic structures in both the intermediate and high field regimes.
\end{abstract}
\maketitle
\section{Introduction}
Binary rare-earth (RE) oxides $A$RE$_2$O$_4$, where $A$ is Ba, Ca, Sr or Eu, have been actively studied for their potential in various applications, particularly in optics and photoluminescence~\cite{Sun_2013,Singh_2016,Kim_2017,Taikar_2018,Zhang_2019}. 
The high density of the materials and large magnetic moments associated with the RE ions also make them promising for low-temperature demagnetisation cooling applications, as magnetocaloric effects are significant~\cite{Midya_2012,Jiang_2018}.
Interest in the fundamental magnetic properties of the $A$RE$_2$O$_4$ oxides was ignited by the realisation that the magnetic RE ions form zigzag ladders and therefore are likely to be governed by geometrically frustrated interactions~\cite{Karunadasa_2005}.
Further studies of several members of the family have confirmed this conjecture and reported the behaviour typically associated with frustrated magnetism, i.e. reduced transition temperatures, partial magnetic ordering, coexistence of different magnetic structures as well as a high sensitivity to an applied field~\cite{Petrenko_2008,Quintero_2012,Young_2013,Aczel_2014,Wen_2015}.

\begin{figure}[tb] 
\centering
\includegraphics[width=0.8\columnwidth]{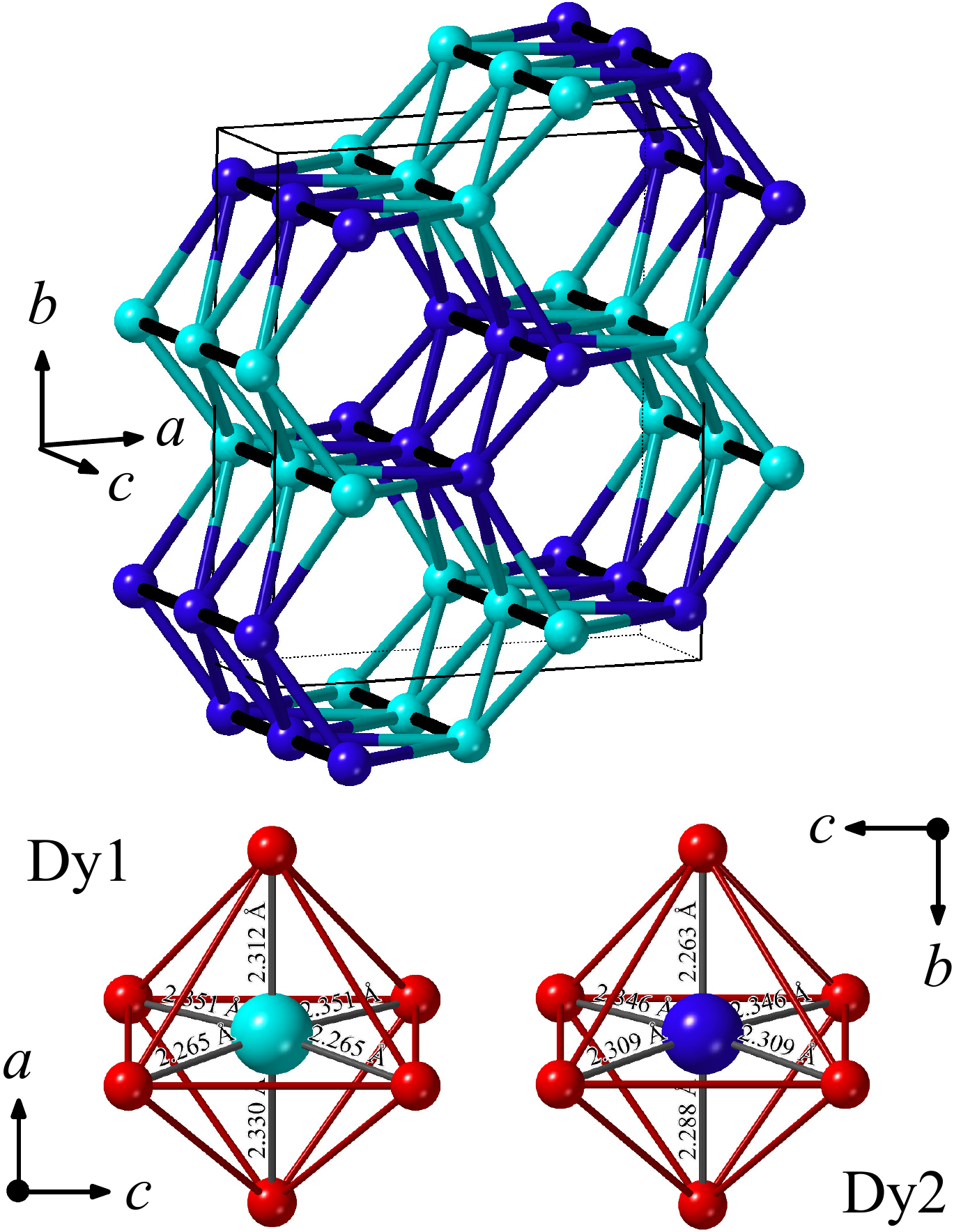}
\caption{Top, positions of the magnetic Dy$^{3+}$ ions within \bdo.
The Dy1 and Dy2 crystallographic positions (space group Pnam) are shown in different colours.
Bottom, DyO$_6$ octahedra for the Dy$^{3+}$ ions in the two crystallographic positions, Dy1 (left) and Dy2 (right), with the distances to the nearest oxygen atoms labeled~\cite{Besara_2014}.}
\label{Fig1_structure}
\end{figure}

One of the family members, \bdo, has recently been reported to order magnetically below $T_{\rm N}=0.48$~K~\cite{Prevost_2018} with the ground state being characterised by the two propagation vectors, \Kone\ and \Ktwo.
This behaviour is in sharp contrast to the properties of \sdo, a homologous compound with very similar unit cell parameters as well as the positions of the magnetic Dy$^{3+}$ ions within the unit cell, see Fig.~\ref{Fig1_structure}~\cite{Karunadasa_2005,Doi_2006,Prevost_2018}.
In zero field, \sdo\ lacks three-dimensional long-range magnetic order down to temperatures as low as 60~mK, its zero-field magnetic structure consists of a collection of antiferromagnetic chains running along the $c$~axis with very little correlations between the chains~\cite{Petrenko_2017}.
It has been argued that one-dimensional trapped domain walls are responsible for the suppression of the 3D order in zero field in this compound~\cite{Gauthier_2017_b}.
However, an {\it up-up-down} (uud) state with much longer correlations can be induced in \sdo\ by a magnetic field applied along the $b$~axis~\cite{Petrenko_2017}.
The field-induced magnetic ordering in \sdo\ for $H \parallel b$ is best seen in single crystal calorimetry~\cite{Cheffings_2013}, ultrasound~\cite{Bidaud_2016} and neutron diffraction~\cite{Petrenko_2017,Gauthier_2017_c} measurements, although, even in a polycrystalline neutron diffraction experiment~\cite{Petrenko_2017}, the powder averaged signal is dominated by the sharp magnetic Bragg peaks formed on top of broad diffuse features seen in zero field.
The uud states are characterised by a 1/3-magnetisation plateau and have also been observed in other compounds in the \sRo\ family, such as \seo\ and \sho~\cite{Hayes_2012}.

Unlike the case of \sdo\ for which large single crystal samples have been available for some time~\cite{Balakrishnan_2009}, \bdo\ has so far only been synthesised in a polycrystalline form or as crystals not exceeding $1.0 \times 0.1 \times 0.1$~mm$^3$ in size~\cite{Besara_2014}, therefore detailed single crystal neutron diffraction studies of \bdo\ remain problematic.
In this paper we report the low-temperature in-field behaviour of \bdo\ investigated through powder neutron diffraction (PND) experiments as well as magnetisation measurements. 
We followed the evolution of the magnetic structure with an applied magnetic field and found that in the intermediate fields, 2.5 to 22.5~kOe, an \afm\ field-induced order is markedly different from the zero-field structure, while the fields above 30~kOe stabilise a polarised \KO\ state, where only one of the Dy$^{3+}$ sites has magnetic moments pointing predominantly along the $b$~axis while the other site remains largely disordered.
Remarkably, the \afm\ field-induced structure characterised by the propagation vector $(0 0 \frac{1}{3})$ appears to be much more robust than the zero-field state, as it survives to temperatures as high as three times $T_{\rm N}(H=0)$.

\section{Experimental procedures}

Polycrystalline samples of \bdo\ were prepared by the conventional solid-state synthesis method following previous reports on preparing this material~\cite{Doi_2006,Besara_2014,Prevost_2018}.
High purity Dy$_2$O$_3$ (99.99\%) and BaCO$_3$ (99.98\%) were used as starting materials.
To ensure the appropriate composition of the final compound, the dysprosium oxide was pre-annealed in air at 1000~$^\circ$C overnight.
A 12.5\% excess of BaCO$_3$ was added to compensate for the loss of barium due to evaporation during the synthesis process.
Powders of the starting materials were weighed, mixed together and pressed into pellets to facilitate the chemical reaction.
The pellets were then transferred into an alumina boat and heated at 1200-1300~$^\circ$C in flowing Ar+3\%H$_2$ gas atmosphere for 4-8 hours~\cite{Doi_2006}.
Powder x-ray diffraction measurements, using a Panalytical X-ray diffractometer with a Cu K$\alpha 1$ anode ($\lambda =1.5406$~\AA), were used to check the composition of the \bdo\ sample prior to the neutron diffraction measurements.

The magnetisation measurements were made in applied magnetic fields of up to 70 kOe using a Quantum Design MPMS SQuID magnetometer.
An iQuantum $^3$He insert~\cite{Shirakawa_2004} allowed the temperature range explored to be extended down to 0.48~K.

PND experiments were conducted on the WISH time-of-flight diffractometer~\cite{WISH} at the ISIS facility at the Rutherford Appleton Laboratory (STFC, UK) in a range of temperatures from 40~mK to 10~K and in fields of up to 60~kOe.
A dilution refrigerator inside a vertical-field cryomagnet (with the opening of $\pm 7.5$~degrees) provided the necessary sample environment.
The sample was placed inside an annular shaped thin-wall copper container in order to minimise high neutron absorption caused mainly by $^{161}$Dy and $^{164}$Dy isotopes in naturally occurring dysprosium.
We were able to add extra He exchange gas to the container during the measurements to ensure full thermal stabilisation of the sample.
This feature of the experimental setup has proved to be crucial in reaching low temperatures for a sample with very limited thermal conductivity.

We used the data obtained in zero field at $T=10$~K as a background to isolate the magnetic signal.
On application of an external magnetic field, a significant redistribution of the grains in the powder sample was observed, as the data taken in a zero field afterwards demonstrate the presence of preferred orientation effects.
This unavoidable complication has resulted in a small ``over-subtraction" for the background signal for some fields and restricted the accuracy of the magnetic Bragg peaks intensity integration. 

\section{Results and discussion}

\subsection{Magnetisation measurements}
\begin{figure}[tb] 
\centering
\includegraphics[width=0.975\columnwidth]{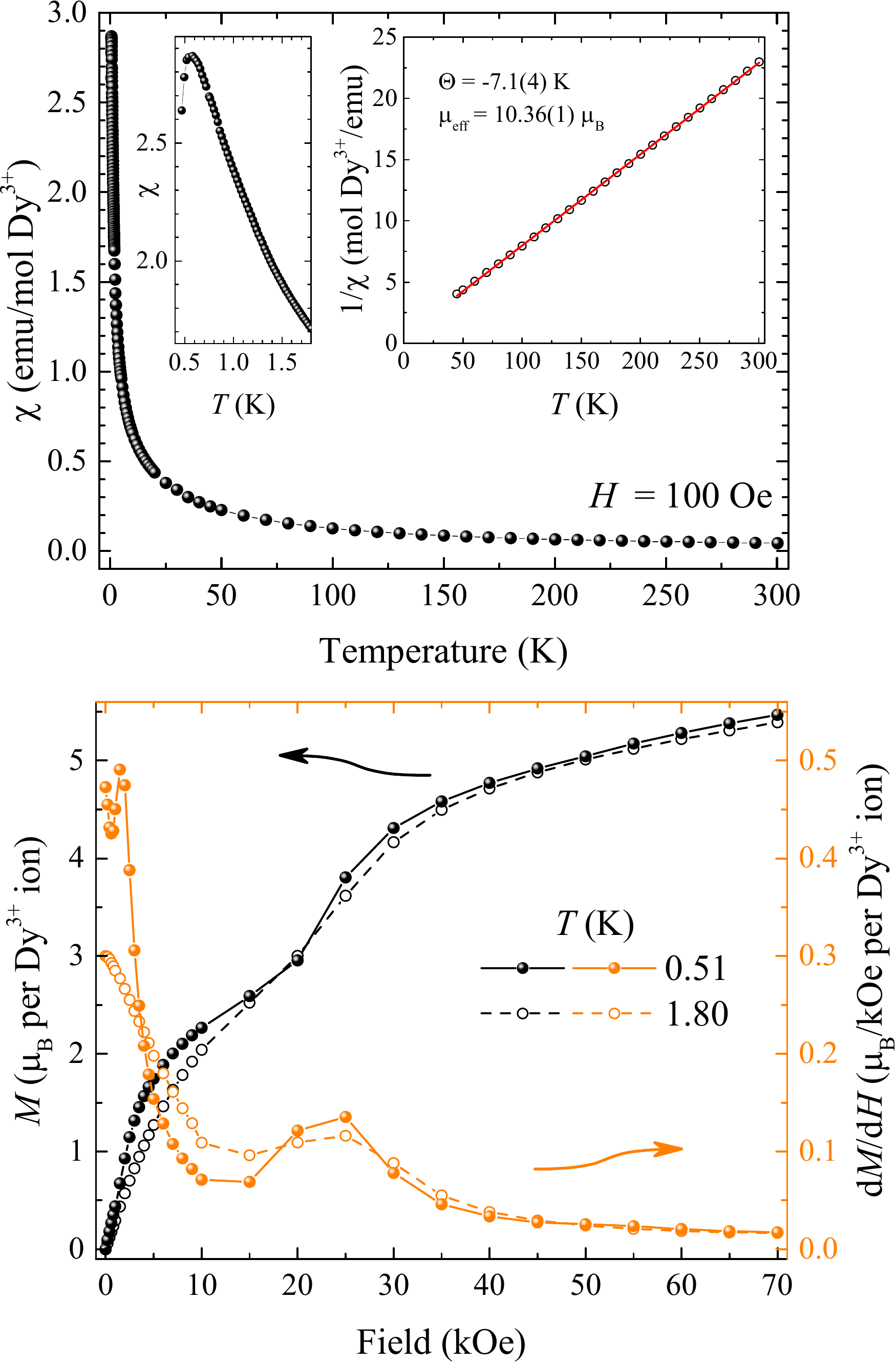}
\caption{(Top) Temperature dependence of the magnetic susceptibility of \bdo\ measured in a field of 100 Oe.
The left inset expands on the low-temperature part of the data, the right inset shows a linear fit to the inverse susceptibility performed in the range 45 to 300~K.
(Bottom) Magnetisation, $M(H)$ and its derivative, $dM/dH$ measured at two different temperatures, 0.51 and 1.80~K.}
\label{Fig2_magnetisation}
\end{figure}

For $T>2$~K, the temperature dependence of magnetic susceptibility, $\chi(T)$, in \bdo\ (see Fig.~\ref{Fig2_magnetisation}) demonstrates a behaviour consistent with the previous observations~\cite{Doi_2006,Besara_2014}.
Although never perfectly linear, the $\chi^{-1}(T)$ largely follows the Curie-Weiss law for  $T>20$~K.
When restricted to the temperature range 45 to 300~K, a linear fit to $\chi^{-1}(T)$ returns the Weiss temperature $\Theta$ of $-7.1(1)$K and the effective magnetic moment $\mu_{\rm eff}$ on the Dy$^{3+}$ ions of $10.36(1)\mu_{\rm B}.$
While the observed magnetic moment is very close to the value calculated using Hund's rules as well as the reported values, $\Theta$ is significantly smaller (in absolute value) compared to -23.5~K from Ref.~\onlinecite{Doi_2006} and to -11.5(4)~K from Ref.~\onlinecite{Besara_2014}.
The difference is most likely due to the different temperature ranges used for the fitting, as well as the presence of ferromagnetic impurities (up to a 2\% level) in the samples prepared using stainless steel crucibles, as in Ref.~\onlinecite{Besara_2014}.

On extending the temperature range for the measurements down to 0.48~K (see the inset in Fig.~\ref{Fig2_magnetisation}) we find that the susceptibility continues to grow at lower temperature until about 0.6~K, where it exhibits a broad maximum before reducing slightly at lower $T$.
The observed broad maximum in $\chi(T)$ is likely to be a precursor for the ordering transition expected at 0.48~K~\cite{Prevost_2018} with the transition temperature itself being marked by a very sharp decrease in $\chi(T)$, as has been seen in powder samples of BaNd$_2$O$_4$~\cite{Aczel_2014} and SrNd$_2$O$_4$~\cite{Riberolles_2019}.

Magnetisation versus field $M(H)$ curves have been measured at 0.51 and 1.80~K. 
At both temperatures there is a plateau (that is more pronounced at lower $T$) from about 5 to 20~kOe (see Fig.~\ref{Fig2_magnetisation}).
The magnetic moment grows continuously in higher fields reaching $5.5\mu_{\rm B}$ per Dy$^{3+}$ ion in 70~kOe, suggesting that in this field the system is far from full saturation.
The value of magnetisation on the plateau, around $2.4\mu_{\rm B}$, amounts to significantly more than a third of the magnetisation measured in 70~kOe, however, this result for a powder-averaged measurement on a sample containing two magnetic sites is not surprising.
In fact the existence of the magnetisation plateaus is a common feature for many \sRo\ compounds, but their presence is much more obvious in the $M(H)$ curves of single crystals~\cite{Hayes_2012} rather than powder samples~\cite{Karunadasa_2005}.
However, even for powder samples the derivative of the magnetisation, $dM/dH$, demonstrates a clear minimum associated with a plateau and in that respect the magnetisation data for \bdo\ (see Fig.~\ref{Fig2_magnetisation}) look remarkably similar to what has been reported for \sdo~\cite{Karunadasa_2005}.

\subsection{Powder neutron diffraction}
\begin{figure}[tb] 
\centering
\includegraphics[width=0.95\columnwidth]{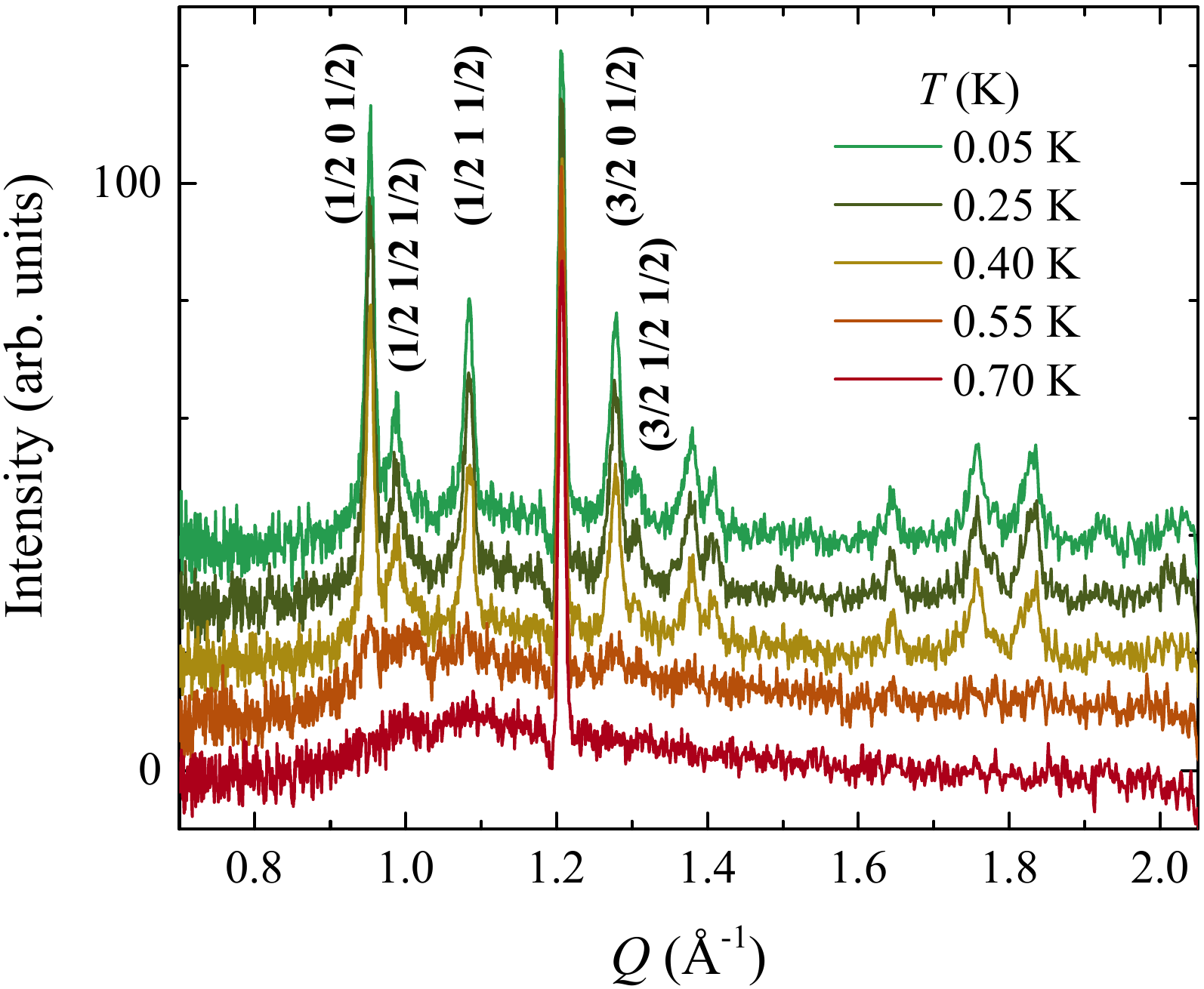}
\caption{Temperature evolution of the magnetic neutron diffraction patterns of \bdo\ in zero applied field.
The magnetic contribution is obtained by subtracting a 10~K background.
The sharp feature at 1.206~\AA$^{-1}$ is an artefact due to redistribution of the grains in the powder sample in an applied field (see text).}
\label{Fig3_zero_field}
\end{figure}

In zero field, powder neutron diffraction measurements on \bdo\ reveal the presence of strong diffuse background signal at higher temperatures and the development of magnetic Bragg peaks at low temperatures, as shown in Fig~\ref{Fig3_zero_field}.
In good agreement with the previously published data~\cite{Prevost_2018}, the peaks are largely confined to temperatures below 0.5~K, as for $T=0.55$~K only remnants of the strongest reflections are present.
The peaks can be indexed using two propagation vectors, \Kone\ and \Ktwo, the indices are given in Fig.~\ref{Fig3_zero_field} for the first five peaks observed at the lowest $Q$.
Note, that the zero field data at higher temperatures were collected after the sample's exposure to high fields at the base temperature.
Therefore in Fig.~\ref{Fig3_zero_field} there is a large nuclear peak at 1.206~\AA$^{-1}$, which appears due to a preferred alignment of the powder grains by the applied field.
As the zero-field magnetic structure has been discussed in Ref.~\cite{Prevost_2018} we do not discuss it in any detail in this paper focusing instead on the in-field behaviour.

\begin{figure}[b] 
\centering
\includegraphics[width=0.95\columnwidth]{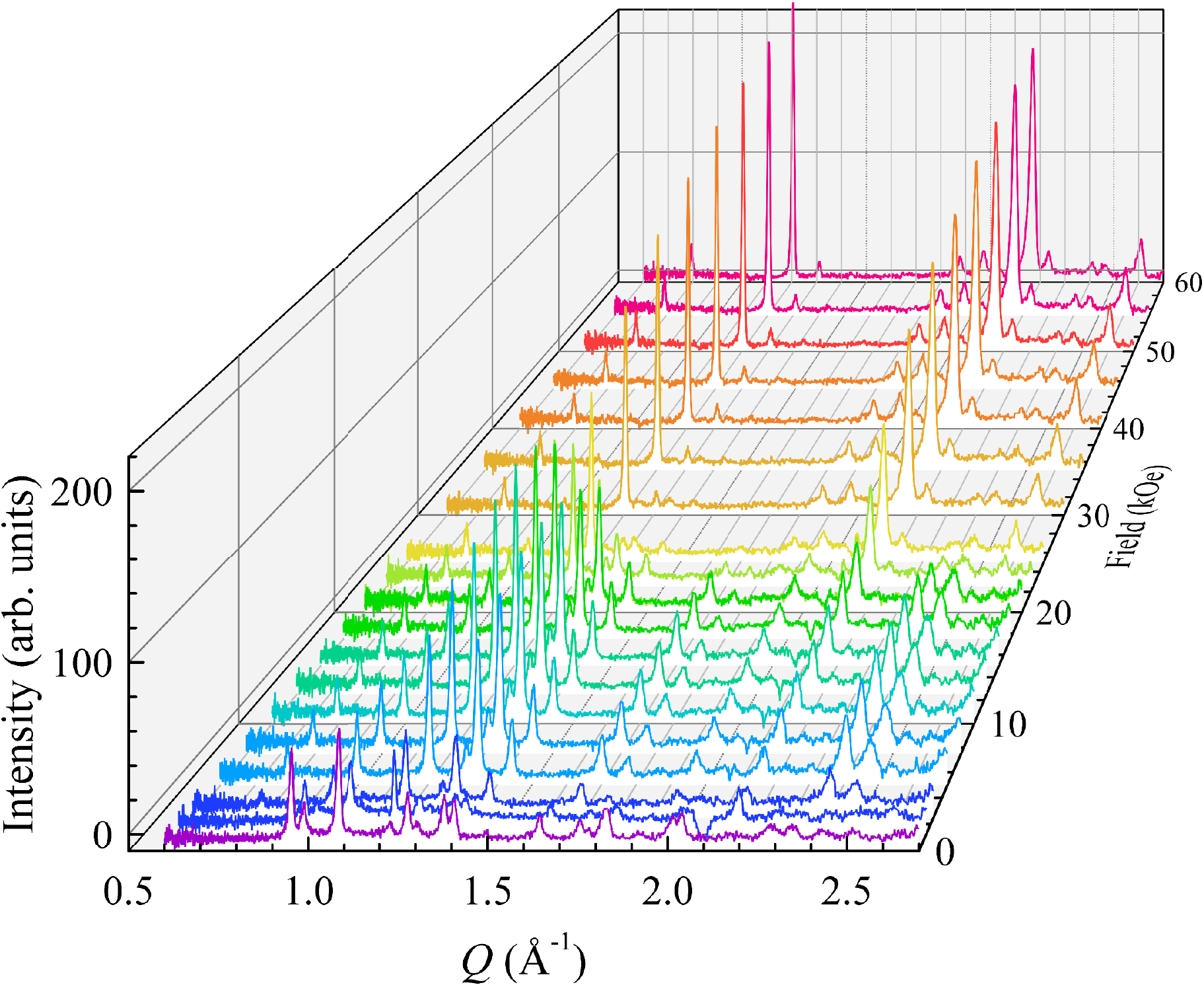}
\caption{Evolution of the magnetic neutron diffraction patterns of \bdo\ with an applied field at $T=0.05$~K in the range 0 to 60~kOe as measured by detector banks 2 and 9 of the WISH diffractometer.
The magnetic contribution is obtained by subtracting a 10~K background.}
\label{Fig4_PND_in_field}
\end{figure}

Figure~\ref{Fig4_PND_in_field} shows the development of the magnetic diffraction patterns of \bdo\ with the applied magnetic field at the base temperature of the cryomagnet, 0.05~K.
We used a 10~K background to isolate the magnetic contribution presuming that the nuclear intensity from the \bdo\ as well as (predominantly copper) intensity from the sample environment remains temperature independent below 10~K.
Although the magnetic signal was seen in all ten detector banks on the WISH diffractometer, we focus mostly on the diffraction patterns from banks 2\&9 and 3\&8 as they provide good $Q$-range coverage combined with high flux and resolution.

The first clear observation from Fig.~\ref{Fig4_PND_in_field}, is that the zero-field \afm\ peaks, the strongest being observed at 0.953 and 1.084~\AA$^{-1}$, start losing their intensity very quickly in an applied field (they are only a half as strong in 1.25~kOe compared to zero field) and by 2.5~kOe they completely disappear.
This extreme sensitivity of the ground state to a weak perturbation is unusual,  but not entirely unexpected -- in zero field, the magnetic structure which appears below 0.48~K seems to have Bragg peaks {\em not} in the positions where diffuse signal is the strongest at $T>T_{\rm N}$, as best seen in the two-dimensional colour plots, Fig.~6a of Ref.~\onlinecite{Prevost_2018}.
It is therefore reasonable to surmise that in zero field, there are several nearly-degenerate states which \bdo\ could adopt and that the ground state has an energy marginally lower than the states influencing the short-range correlations just above the $T_{\rm N}$.
Given this peculiar behaviour, it might be of interest to revisit the zero-field diffraction data obtained above the ordering temperature and to employ reverse Monte Carlo refinements to study the magnetic correlations~\cite{Paddison_2012,Paddison_2013}.
 
 \begin{figure}[tb] 
\centering
\includegraphics[width=0.975\columnwidth]{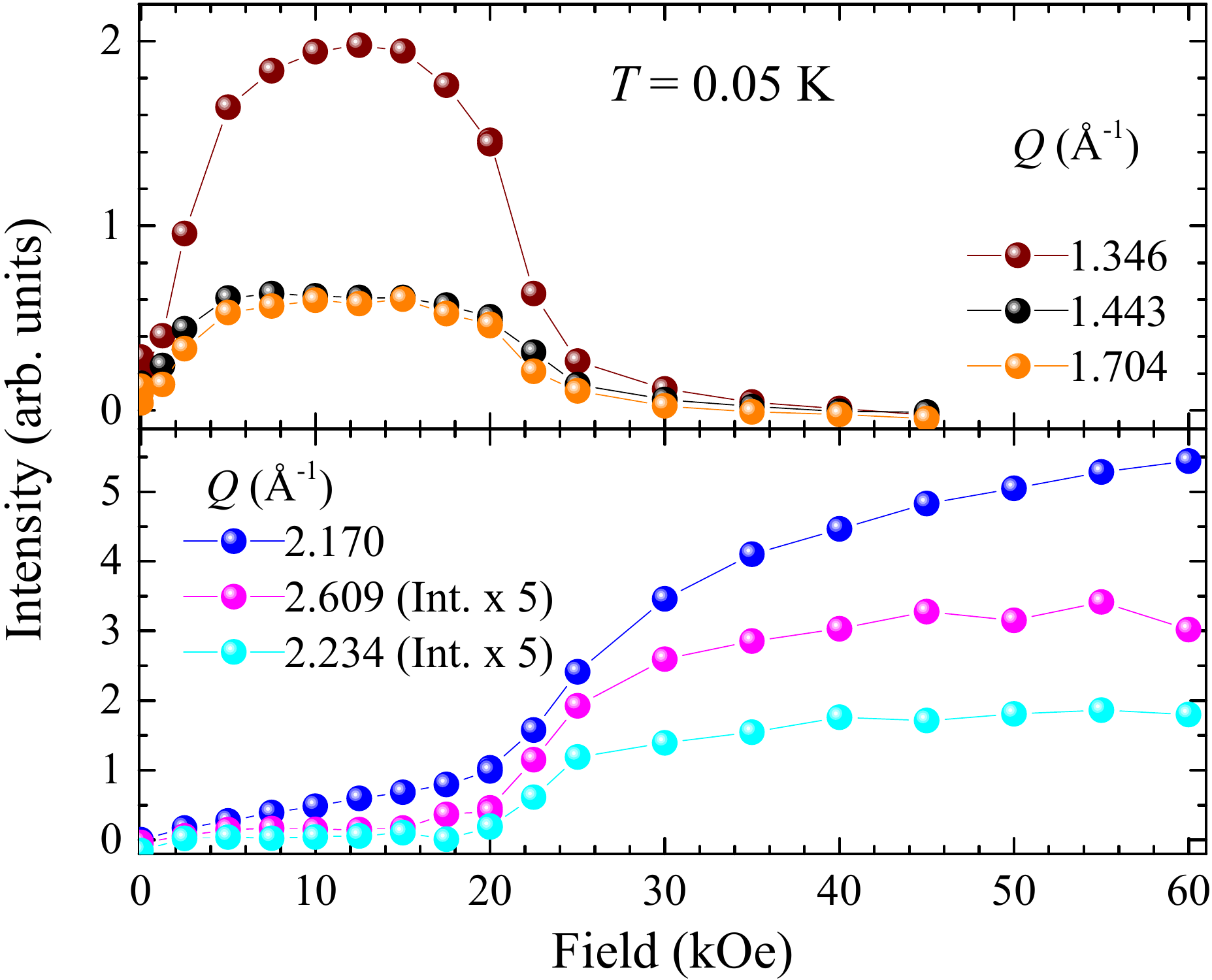}
\vspace{-1mm}
\caption{Field dependence of the intensity for the (top panel) AFM and (bottom panel) FM peaks in \bdo.
The peaks are labeled by their scattering vector positions.
The intensity of the peaks at 2.234 and 2.609~\AA\ is increased by a factor of 5.               
The intensity is obtained by integrating the peaks above zero after subtracting a 10~K background signal.}
\label{Fig5_Int}
\end{figure}

Instead of the zero-field peaks the applied field induces two other sets of magnetic Bragg peaks.
The intensities of the first set of reflections appear in the intermediate applied fields and then vanish rather rapidly in higher applied fields, becoming negligibly small in fields above 25~kOe, we label these peaks as \afm, AFM.
Another set of peaks gain some intensity in intermediate fields, continue to increase in intensity in higher fields, show a tendency to saturation, but do not fully saturate in the highest applied field of 60~kOe.
We label these peaks as ferromagnetic, FM.
This drastically different behaviour of the two sets of peaks is emphasised in Fig.~\ref{Fig5_Int} which shows the field dependence of the intensity of a selected number of the AFM and FM peaks.
Fig.~\ref{Fig5_Int} also provides a natural way to introduce more accurately the definitions of the intermediate and high field regions, the former being from 2.5 to 25~kOe and the latter above 25~kOe.
The intermediate fields region coincides with the stabilisation of a magnetisation plateau.

\begin{figure}[tb] 
\centering
\includegraphics[width=0.9\columnwidth]{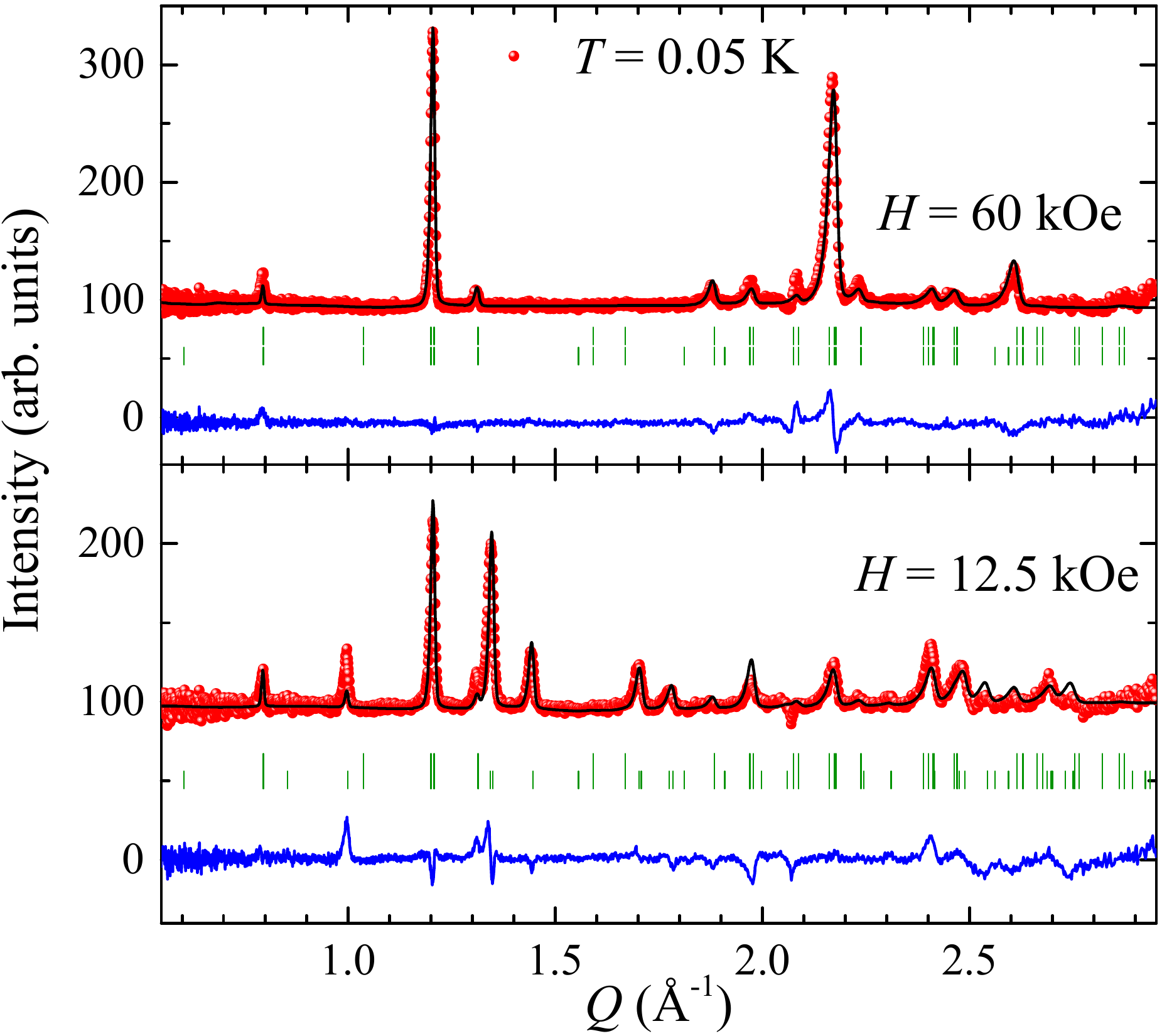}
\caption{Rietveld refinement of the PND data in an applied magnetic field of (top panel) 60~kOe and (bottom panel) 12.5~kOe at 0.05~K.
The experimental data (red circles) from banks 2 and 9 of the WISH diffractometer~\cite{WISH} after subtracting a 10~K background signal, calculated patterns (black lines) and the difference curves (blue lines) are shown.
Ticks represent the positions of the nuclear and magnetic Bragg peaks.}
\label{Fig6_refinement}
\end{figure}

Despite the natural limitations of magnetic structure determination from PND in an applied magnetic field, we were able to refine the diffraction data both in the FM (60~kOe) and the AFM (12.5~kOe) regimes, see Fig.~\ref{Fig6_refinement}.

\begin{figure}[tb] 
\centering
\includegraphics[width=0.95\columnwidth]{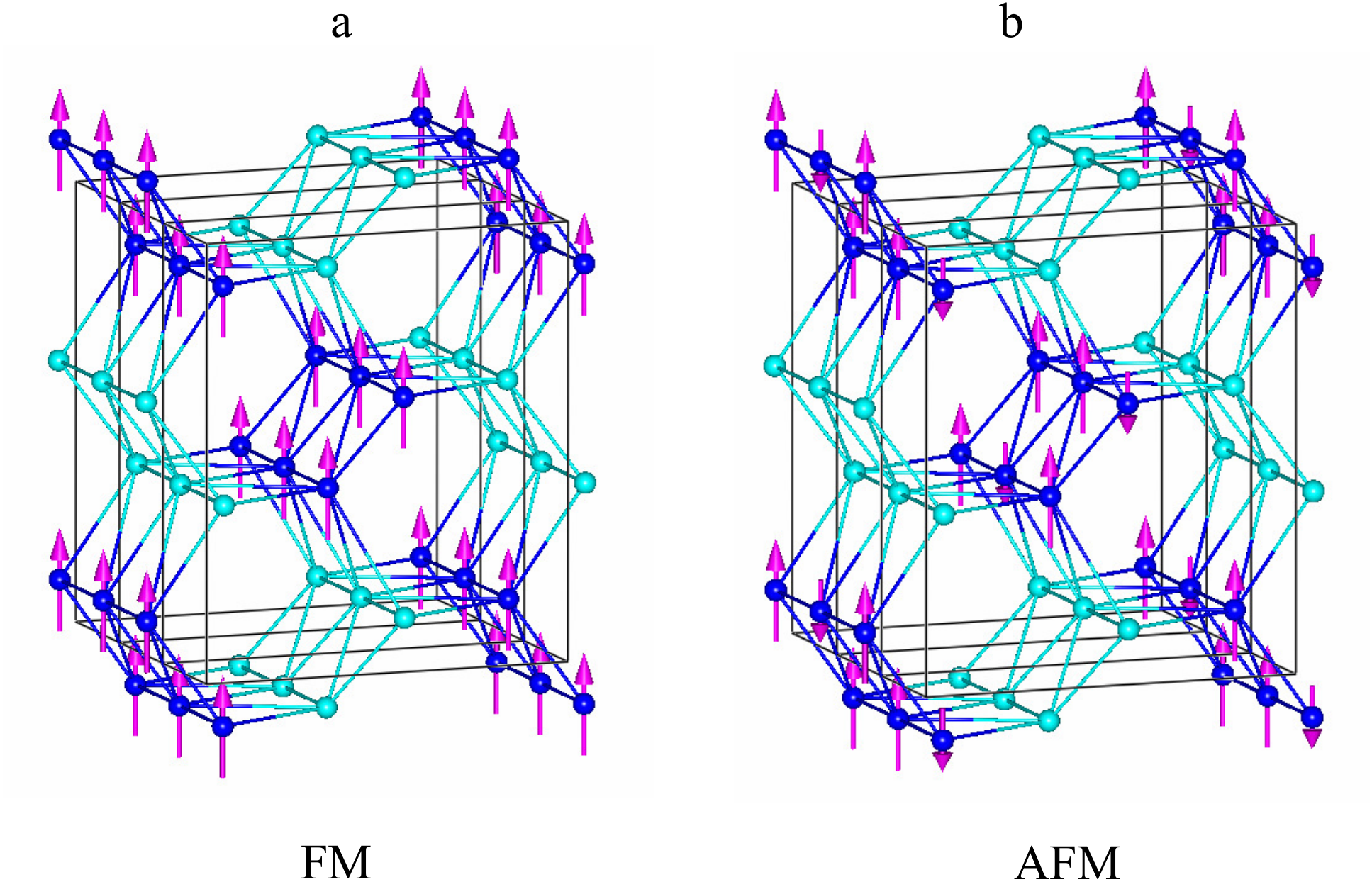}
\caption{Magnetic structures of \bdo\ in the (a) FM and (b) AFM regimes as obtained from PND data refinements at $T=0.05$~K.}
\label{Fig7_structures}
\end{figure}

The high-field data (60~kOe) can be successfully fitted assuming a simple ferromagnetic ordering with the moments aligned along the $b$~axis.
The refinement yields the ordering only on one Dy-site with the ordered moment of $9.44(5)\mu_{\rm B}$, and no statistically significant ordered component along the $a$, $b$ or $c$ directions was detected for the second Dy site.
Attempts to constrain the moment size of Dy1 and Dy2 to be equal failed to reproduce the experimental diffraction pattern.
It  is impossible to determine from the powder diffraction data which of the Dy$^{3+}$ ions on crystallographically inequivalent sites participate in the formation of the FM state and which stays disordered, as both possibilities provide equally good fits to the data.
However, the crystal electric field calculations reported in Ref.~\cite{Prevost_2018} indicate that the Dy1 and Dy2 sites have the easy axes parallel to the $a$ and $b$ directions respectively.
The Dy2 site also has a very large separation between the ground state and the first excited doublet~\cite{Prevost_2018} and therefore is likely to possess an almost perfect Ising-like magnetic moment.
This observation provides a good reason to locate the ordered moment of the high-field phase of \bdo\ on the Dy2 site, as shown in Fig.~\ref{Fig7_structures}a.

Apparently, the presence of only one ordered site is closely related to the substantial preferred orientation of the powder particles induced by the applied magnetic field.
The refinement reveals that a significant fraction of the sample is textured with the preferred orientation vector $(0,1,0)$ lying along the direction of the applied field.

A similar situation, where only one site is ordered and the sample is substantially textured has been found in the intermediate field regime, 2.5 to 25~kOe.
For this field region the magnetic peaks are indexed with two propagation vectors, ${\bf k}=0$ and ${\bf k}^\prime =[0~0~\frac{1}{3}]$.
These coexisting propagation vectors strongly point to the uud type of the magnetic structure.
Indeed, the model where magnetic phases of the four symmetry related Dy sites in the 4c Wyckoff position differ by $\frac{2\pi}{3}$ provides a reasonably good quality fit for the experimentally observed intensities (Fig.~\ref{Fig6_refinement}).
The corresponding uud magnetic structure with moments polarised along the $b$~axis is shown in Fig.~\ref{Fig7_structures}b.
The ordered moment for the up and down spins are $9.2(1)\mu_{\rm B}$  and $5.8(1)\mu_{\rm B}$ respectively, refined for the diffraction pattern collected at 12.5~kOe.
Again, it was not possible to decide which Dy1 or Dy2 carries the ordered moments based on the refinement quality, but the moment direction strongly suggests the Dy2 site. 

\begin{figure}[tb] 
\centering
\includegraphics[width=0.995\columnwidth]{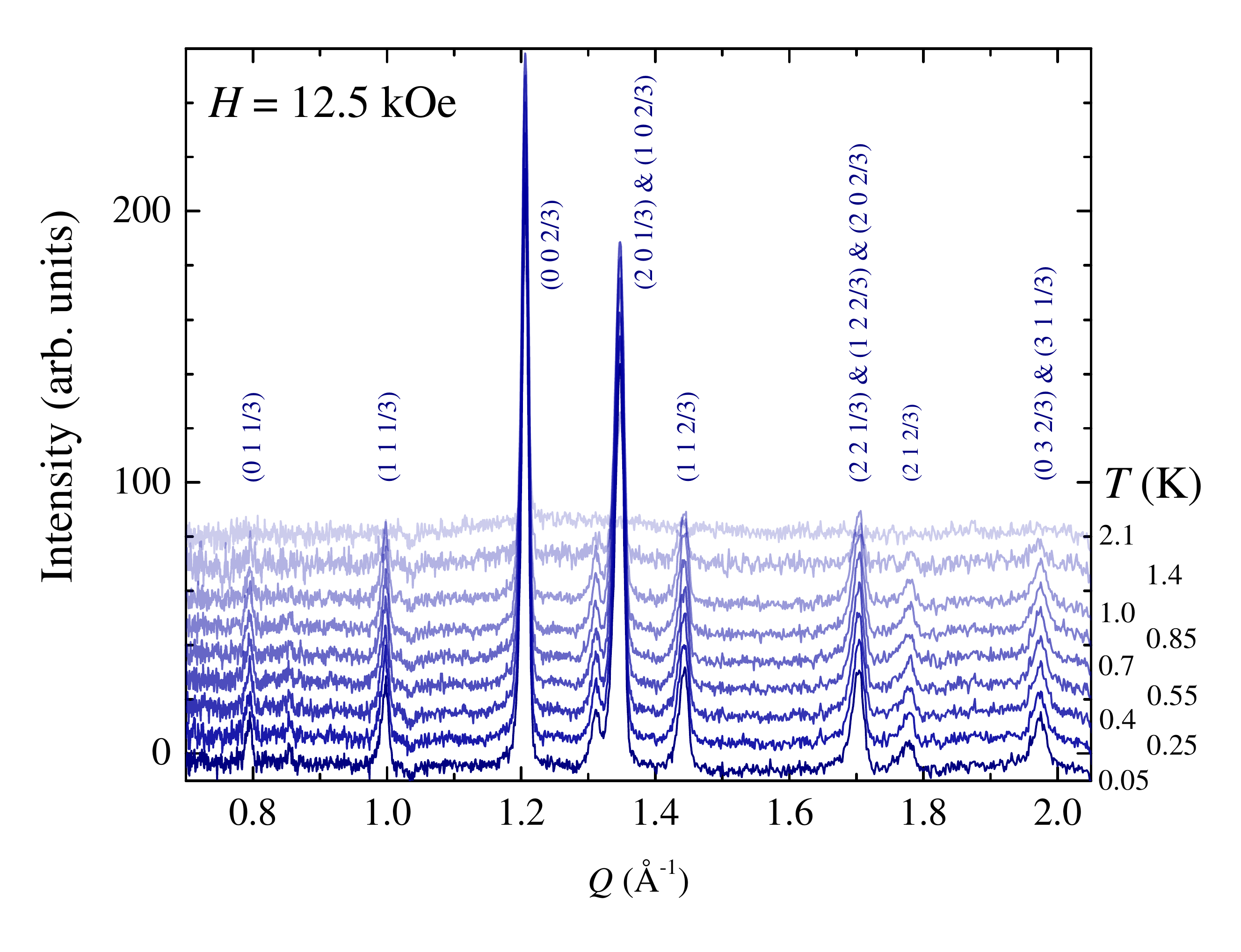}
\caption{Temperature evolution of the magnetic diffraction patterns of \bdo\ in 12.5~kOe.
The magnetic contribution is obtained but subtracting the 10~K zero-field background.
The curves are consecutively offset by 10 units for clarity.}
\label{Fig8_AFM_temp}
\end{figure}

Fig.~\ref{Fig8_AFM_temp} shows the evolution of the diffraction patterns observed in an intermediate field of 12.5~kOe with increasing temperature.
It is rather clear that none of the AFM peaks are greatly affected by the temperature increase until well above $T_{\rm N}(H=0)$.
In fact they show almost no temperature dependence for temperatures below 1~K, survive at temperature as high as 1.4~K and disappear completely only at $T=2.1$~K.
We have also collected diffraction data in a field of 5~kOe (not shown) in increasing temperature, which support the observation that the AFM phase survives at temperature above 1.4~K.
Note, the rather sparse temperature points at higher temperatures is due to the limitations imposed by the dilution refrigerator used, as it cannot maintain temperatures above 1~K for time sufficiently long to collect data, while the base temperature of the cryomagnet without dilution refrigerator cooling was 2.1~K.

As stated above, the appearance of the magnetisation plateaux is a common feature of many similar \sRo\ compounds~\cite{Hayes_2012,Petrenko_2014}.
What makes the \bdo\ case unusual is the fact that the field-induced uud state in this compound seems to be completely different from the zero-field structure rather than evolving from it.
In \sho, for example~\cite{Young_2019}, the main effect of a field applied along the $b$~axis is in the evolution of the diffuse magnetic scattering to a set of nearly resolution-limited Bragg peaks associated with the uud state.
Similarly in \seo, a field applied parallel to the $a$~axis drives the planes of diffuse scattering observed in zero field to much more localised, but still diffuse features, corresponding to the establishment of the uud structure and the associated one-third magnetisation plateau~\cite{Riberolles_2020}.
In the absence of single crystal samples of \bdo\ large enough for neutron diffraction, however, it will be extremely difficult check if in this compound the field-induced uud state also has a diffuse precursor in zero field.

Given an unusual robustness of the field-induced AFM state it might also be of interest to perform further magnetocaloric effect measurements on \bdo\ with a view to using it as a cryogenic refrigerant material~\cite{Midya_2012,Jiang_2018}.

\section{Conclusions}
In conclusion, we have investigated the in-field behaviour of \bdo\ at low temperatures and found that the zero-field state is very sensitive to the external fields as it is removed by the application of 2.5~kOe.
We also find that in this zigzag ladder compound an uud structure, corresponding to a magnetisation plateau is induced by the fields between 2.5 and 25~kO, while a polarised FM state is supported in fields above 25~kOe.
Remarkably, the field-induced uud structure is much more stable in higher temperatures compared to the zero-field state, as it  survives warming to the three times higher temperature than the $T_{\rm N}(H=0)$.
We hope that this experimental discovery will stimulate further theoretical developments for the ground state selection in frustrated systems, particularly in an applied field.

\section*{ACKNOWLEDGMENTS}
We are thankful to Stepan Marek for his contribution towards preparation of \bdo\ samples and their characterisation as well as to Martin Lees and Olga Young for careful reading of the manuscript.
The work and dedication of the sample environment team at ISIS is gratefully acknowledged.

\bibliography{BaLn2O4_all}
\end{document}